\documentstyle[12pt,axodraw,epsf]{article}

\catcode`\@=11
\long\def\@makefntext#1{
\protect\noindent \hbox to 3.2pt {\hskip-.9pt  
$^{{\ninerm\@thefnmark}}$\hfil}#1\hfill}                

\def\@makefnmark{\hbox to 0pt{$^{\@thefnmark}$\hss}}  
        
\def\ps@myheadings{\let\@mkboth\@gobbletwo
\def\@oddhead{\hbox{}
\rightmark\hfil\ninerm\thepage}   
\def\@oddfoot{}\def\@evenhead{\ninerm\thepage\hfil
\leftmark\hbox{}}\def\@evenfoot{}
\def\sectionmark##1{}\def\subsectionmark##1{}}

\renewcommand{\thefootnote}{\fnsymbol{footnote}}

\newcounter{sectionc}\newcounter{subsectionc}\newcounter{subsubsectionc}
\renewcommand{\section}[1] {\vspace*{0.6cm}\addtocounter{sectionc}{1} 
\setcounter{subsectionc}{0}\setcounter{subsubsectionc}{0}\noindent 
        {\normalsize\bf\thesectionc. #1}\par\vspace*{0.4cm}}
\renewcommand{\subsection}[1] {\vspace*{0.6cm}\addtocounter{subsectionc}{1} 
        \setcounter{subsubsectionc}{0}\noindent 
        {\normalsize\it\thesectionc.\thesubsectionc. #1}\par\vspace*{0.4cm}}
\renewcommand{\subsubsection}[1]
{\vspace*{0.6cm}\addtocounter{subsubsectionc}{1}
        \noindent {\normalsize\rm\thesectionc.\thesubsectionc.\thesubsubsectionc. 
        #1}\par\vspace*{0.4cm}}

\newcounter{appendixc}
\newcounter{subappendixc}[appendixc]
\newcounter{subsubappendixc}[subappendixc]

\renewcommand{\appendix}[1] {\vspace*{0.6cm}
        \refstepcounter{appendixc}
        \setcounter{figure}{0}
        \setcounter{table}{0}
        \setcounter{equation}{0}
        \renewcommand{\thefigure}{\Alph{appendixc}.\arabic{figure}}
        \renewcommand{\thetable}{\Alph{appendixc}.\arabic{table}}
        \renewcommand{\theappendixc}{\Alph{appendixc}}
        \renewcommand{\theequation}{\Alph{appendixc}.\arabic{equation}}
        \noindent{\bf Appendix \theappendixc #1}\par\vspace*{0.4cm}}

\def\abstracts#1{{
        \centering{\begin{minipage}{12.2truecm}\footnotesize\baselineskip=12pt\noindent
        \centerline{\footnotesize ABSTRACT}\vspace*{0.3cm}
        \parindent=0pt #1
        \end{minipage}}\par}} 

\renewenvironment{thebibliography}[1]
        {\begin{list}{\arabic{enumi}.}
        {\usecounter{enumi}\setlength{\parsep}{0pt}
\setlength{\leftmargin 1.25cm}{\rightmargin 0pt}
         \setlength{\itemsep}{0pt} \settowidth
        {\labelwidth}{#1.}\sloppy}}{\end{list}}

\newcounter{itemlistc}
\newcounter{romanlistc}
\newcounter{alphlistc}
\newcounter{arabiclistc}

\newcommand{\fcaption}[1]{
        \refstepcounter{figure}
        \setbox\@tempboxa = \hbox{\footnotesize Fig.~\thefigure. #1}
        \ifdim \wd\@tempboxa > 6in
           {\begin{center}
        \parbox{6in}{\footnotesize\baselineskip=12pt Fig.~\thefigure. #1}
            \end{center}}
        \else
             {\begin{center}
             {\footnotesize Fig.~\thefigure. #1}
              \end{center}}
        \fi}

\newcommand{\tcaption}[1]{
        \refstepcounter{table}
        \setbox\@tempboxa = \hbox{\footnotesize Table~\thetable. #1}
        \ifdim \wd\@tempboxa > 6in
           {\begin{center}
        \parbox{6in}{\footnotesize\baselineskip=12pt Table~\thetable. #1}
            \end{center}}
        \else
             {\begin{center}
             {\footnotesize Table~\thetable. #1}
              \end{center}}
        \fi}

\def\@citex[#1]#2{\if@filesw\immediate\write\@auxout
        {\string\citation{#2}}\fi
\def\@citea{}\@cite{\@for\@citeb:=#2\do
        {\@citea\def\@citea{,}\@ifundefined
        {b@\@citeb}{{\bf ?}\@warning
        {Citation `\@citeb' on page \thepage \space undefined}}
        {\csname b@\@citeb\endcsname}}}{#1}}

\newif\if@cghi
\def\cite{\@cghitrue\@ifnextchar [{\@tempswatrue
        \@citex}{\@tempswafalse\@citex[]}}
\def\citelow{\@cghifalse\@ifnextchar [{\@tempswatrue
        \@citex}{\@tempswafalse\@citex[]}}
\def\@cite#1#2{{$\null^{#1}$\if@tempswa\typeout
        {IJCGA warning: optional citation argument 
        ignored: `#2'} \fi}}

 1
 1
 1

\font\ninerm=cmr9

\begin{document}

\begin{flushright}
MPI/PhT/97-007\\
hep-ph/9704227
\end{flushright}

\centerline{\normalsize\bf RESONANT CP VIOLATION IN HIGGS PRODUCTION,}
\baselineskip=22pt
\centerline{\normalsize\bf MIXING AND DECAY}
\vspace*{0.2cm}
\centerline{\footnotesize APOSTOLOS PILAFTSIS}\footnote[1]{
To appear in the Proc.\ of the Ringberg Workshop on ``The Higgs puzzle - 
What can we learn from LEP2, LHC, NLC, and FMC?'' ed.\ B.A. Kniehl,
(Tegernsee, 8 -- 13 December 1996, World Scientific, Singapore)}
\baselineskip=13pt
\centerline{\footnotesize\it Max-Planck-Institute f\"ur Physik,
F\"ohringer Ring 6, 80805 Munich, Germany}

\vspace*{0.35cm}
\abstracts{
CP violation induced by resonant transitions of a scalar to a pseudoscalar
intermediate state is analyzed within a gauge-invariant resummation approach
based on the pinch technique. Necessary conditions governing the resonant 
enhancement of CP violation in transition amplitudes are derived. The results 
of this study are then applied to describe the indirect mixing of a CP-even 
Higgs scalar, $H$, with a CP-odd Higgs particle, $A$, in two-Higgs doublet 
models. CP violation through $HA$ mixing in high-energy $pp$, $e^+e^-$ and 
$\mu^+\mu^-$ scatterings are estimated to be very large for a natural choice
of kinematic parameters. Low-energy constraints originating from 
experimental upper bounds on the neutron electric dipole model are implemented
in this analysis.}

\normalsize\baselineskip=15pt
\setcounter{footnote}{0}
\renewcommand{\thefootnote}{\alph{footnote}}
\section{Introduction}

CP asymmetries induced by finite widths of unstable 
particles\cite{AP,PNsusy,CPetal} have received much attention over the last 
few years.  In many new-physics scenarios of the Standard Model (SM) with 
extended fermionic and/or Higgs sector, CP-violating effects may arise
from the interference of the top-quark width with that of a new up-type 
quark $t'$\cite{AP} or from interference width effects between the $W$ 
boson and a new $H^+$ scalar in the partially integrated top decay 
rate.\cite{CPetal} 

Recently, resonant CP-violating transitions of a CP-even Higgs scalar, $H$, 
into a CP-odd Higgs particle, $A$, are studied within the context of 
two-Higgs doublet models.\cite{APRL}  The size of CP violation has been 
estimated to be fairly large, {\em i.e.}, of order one for some choice of 
kinematic parameters. Here, we shall show that these CP-violating phenomena 
have dynamical features very similar to those of the $K^0\bar{K}^0$ 
system.\cite{K0K0} 

The talk is organized as follows: We first derive the necessary
conditions for resonantly enhanced CP violation and so demonstrate explicitly 
that this CP violation induced by $HA$ mixing has common features with the
CP-violating phenomenon through indirect mixing in the kaon system in
Section 2. Bounds from electric dipole moments (EDM's) of the neutron
and electron are discussed in Section 3.  In Section 4, models with 
non-vanishing $HA$ mixing are considered and the physics potential to 
observe resonant CP violation through such a mixing at future $pp$, $e^+e^-$ 
or $\mu^+\mu^-$ machines is briefly discussed. Our conclusions are drawn in 
Section 5.

\section{Necessary conditions for resonant CP violation}

To deduce the necessary conditions under which CP violation can
be resonantly enhanced in the $HA$ system, we shall perform our
analysis in a $K^0\bar{K}^0$-like basis.\cite{Misha}  To this end, 
the relation between the $K^0\bar{K}^0$ and $HA$ bases may be given by
the following transformations:
\begin{eqnarray}
\label{KKtrafo}
iA &=& \frac{1}{\sqrt{2}}\, \Big(\, K^0\, -\, \bar{K}^0\, \Big)\,
,\nonumber\\ H &=& \frac{1}{\sqrt{2}}\, \Big(\, K^0\, +\, \bar{K}^0\,
\Big)\, ,
\end{eqnarray}
where $\bar{K}^0$ is the Hermitian and CP-conjugate state of
$K^0$. Expressing the effective Hamiltonian ${\cal H}(s)$ in the 
$K^0\bar{K}^0$ basis, we get
\begin{eqnarray}
\label{KKHam}
(K^{0*},\bar{K}^{0*})\tilde{{\cal H}}\left(\begin{array}{c} K^0\\
\bar{K}^0
\end{array}\right) & = & \nonumber\\
&&\hspace{-125pt} \frac{1}{2}\, \left[\begin{array}{cc}
M^2_H+M^2_A-\widehat{\Pi}^{HH}-\widehat{\Pi}^{AA} &
M^2_H-M^2_A-\widehat{\Pi}^{HH}+\widehat{\Pi}^{AA}+2i\widehat{\Pi}^{AH}\\
M^2_H-M^2_A-\widehat{\Pi}^{HH}+\widehat{\Pi}^{AA}-2i\widehat{\Pi}^{AH}
& M^2_H+M^2_A-\widehat{\Pi}^{HH}-\widehat{\Pi}^{AA}
\end{array} \right] ,\quad
\end{eqnarray}
where the hat on the $HH$, $AA$ and $HA$ self-energies has two meanings.
First, it denotes that the diagonal self-energies are renormalized in 
some natural scheme, {\em e.g.}, on-shell renormalization scheme, whereas
the CP-violating $HA$ vacuum amplitudes are ultra-violet (UV) finite
in our models and do not require renormalization. Second, the symbol hat 
indicates that the self-energies are evaluated within a gauge-invariant
resummation based on the PT.\cite{JP&AP,AS} It is now easy to observe that 
$\tilde{{\cal H}}$ in Eq.\ (\ref{KKHam}) possesses all known properties of 
the kaon system. In particular, CPT demands 
\begin{equation}
\label{CPTHam}
\tilde{{\cal H}}_{11}(s)\ =\ \tilde{{\cal H}}_{22}(s)\, ,
\end{equation}
which is valid in Eq.\ (\ref{KKHam}), while CP invariance is reassured only
if
\begin{equation}
\tilde{{\cal H}}_{12}(s)\ =\ \tilde{{\cal H}}_{21}(s)\, .
\end{equation}
The latter can only be true if $\widehat{\Pi}^{AH}(s)=0$. Consequently, the 
presence of a non-zero $HA$ mixing leads to CP violation in the effective
Hamiltonian $\tilde{{\cal H}}(s)$. In particular, it is known that the basic 
parameter measuring CP violation through indirect mixing in the kaon 
system\cite{K0K0} is given by
\begin{eqnarray}
\label{q/p}
\Big| \frac{q}{p}\Big|^2 &=& \left|\frac{\tilde{{\cal H}}_{21}}{
\tilde{{\cal H}}_{12}}\right| \nonumber\\ &=& \left\{ \frac{
[M^2_H-M^2_A-2\Im m\widehat{\Pi}^{HA}]^2+ [\Im
m(\widehat{\Pi}^{HH}-\widehat{\Pi}^{AA})+2\Re e\widehat{\Pi}^{HA}]^2}{
[M^2_H-M^2_A+2\Im m\widehat{\Pi}^{HA}]^2+ [\Im
m(\widehat{\Pi}^{HH}-\widehat{\Pi}^{AA})-2\Re e\widehat{\Pi}^{HA}]^2}
\right\}^{1/2}.
\end{eqnarray}
It will prove useful to consider the following two cases:

\begin{itemize}
\item[ (i)] The $HA$ mixing occurs at the tree level or is
induced radiatively after integrating out heavy degrees of freedom,
{\em i.e.}, $\Re e\widehat{\Pi}^{HA}\not= 0$ and it is UV
safe. We also assume $\Im m \widehat{\Pi}^{HA} = 0$. Then, for 
$M_H\approx M_A$, the CP-violating mixing parameter behaves as
\begin{equation}
\label{q/p1}
\Big| \frac{q}{p}\Big|^2\ \sim \ \left| \frac{ \Im
m(\widehat{\Pi}^{HH}-\widehat{\Pi}^{AA})+2 \Re
e\widehat{\Pi}^{HA}}{\Im m(\widehat{\Pi}^{HH}-\widehat{\Pi}^{AA})-
2\Re e\widehat{\Pi}^{HA}}\right|\ .
\end{equation}
Clearly, for
\begin{equation}
\label{CPN1}
\Im m(\widehat{\Pi}^{HH}-\widehat{\Pi}^{AA})\ \sim \ \pm\, \frac{1}{2}
\Re e\widehat{\Pi}^{HA}\, ,
\end{equation}
$|q/p|$ takes either very small or very large values, yielding
resonant enhancement of CP violation. For large mass differences,
$M_H-M_A\gg M_H,M_A$, $|q/p|\approx 1$ as can be seen from Eq.\ (\ref{q/p}).

\item[(ii)] Another interesting case arises when $\Im m \widehat{\Pi}^{HA}\not
= 0$ and $\Re e\widehat{\Pi}^{HA} = 0$. For small width differences, 
$\Im m\widehat{\Pi}^{HH}\approx \Im m\widehat{\Pi}^{AA}$, we have
\begin{equation}
\label{q/p2}
\Big| \frac{q}{p}\Big|^2\ \sim \ \left| \frac{M^2_H - M^2_A - 2\Im
m\widehat{\Pi}^{HA}}{ M^2_H - M^2_A + 2\Im m\widehat{\Pi}^{HA}}\right|
\ ,
\end{equation}
giving rise to the condition for resonant CP violation 
\begin{equation}
\label{CPN2}
M^2_H - M^2_A\ \sim \ \pm\, \frac{1}{2} \Im m\widehat{\Pi}^{HA}\, .
\end{equation}

\end{itemize}

It is now important to remark that for maximal CP violation, {\em i.e.}, of
order unity, $|q/p|$ should either vanish or tend to infinity. This implies 
that either $\tilde{{\cal H}}_{12}=0$ or $\tilde{{\cal H}}_{21}=0$, but 
{\em not } both. This limiting case reflects the fact that the two (no-free) 
particles, $H$ and $A$, are exactly degenerate, {\em i.e.}, 
$\overline{M}_H = \overline{M}_A$ and $\overline{\Gamma}_H = 
\overline{\Gamma}_A$, where $\overline{M}^2_{H,A} - i\overline{M}_{H,A}
\overline{\Gamma}_{H,A}$ are the two complex pole-mass eigenvalues of the 
effective Hamiltonian $\tilde{{\cal H}}(s)$ in Eq.\ (\ref{KKHam}).

\section{Constraints from electric dipole moments}

The presence of a $HA$ operator may also contribute to other
low-energy CP-violating observables. CP-violating quantities sensitive
to $HA$ terms is the EDM of the neutron and the electron. The one-loop 
contribution of the $HA$ mixing to the EDM may be estimated by
\begin{equation}
\label{EDMf}
d_q/e\ \approx\ -Q_q \frac{\alpha_w}{4\pi}\, \frac{m_f}{M^2}\,
\frac{m^2_f}{M^2_W}\, \xi_{HA}\,
\ln\Big(\frac{m^2_f}{M^2}\Big)\, ,
\end{equation}
with $M=(M_H +M_A)/2$, $Q_q$ denoting the fractional charge of the quark
\begin{equation}
\label{xiHA}
\xi_{HA}\ =\ \chi^f_A\chi^f_H\, \frac{\widehat{\Pi}^{HA}(M^2)}{M^2}\ .
\end{equation}
\begin{center}
\begin{picture}(360,160)(0,0)
\SetWidth{0.8}

\ArrowLine(110,30)(145,60)\Text(130,35)[lb]{$q_L$}
\DashLine(145,60)(180,100){5}\Text(146,68)[rb]{$A$}
\Text(162,86)[rb]{$H$}\Text(160,77)[]{{\boldmath $\times$}}
\Photon(180,100)(215,60){3}{5}\Text(200,85)[l]{$\gamma$}
\ArrowLine(145,60)(215,60)\Text(180,51)[l]{$q_R$}
\ArrowLine(215,60)(250,30)\Text(218,35)[lb]{$q_R$}
\Photon(180,100)(180,150){3}{5}\Text(186,140)[l]{$\gamma$}
\GCirc(180,100){10}{0.5}

\end{picture}\\[-0.5cm]
{\small {\bf Fig.\ 1:} The Barr-Zee mechanism for generating
EDM.}
\end{center}
\vspace*{0.2cm}

The experimental upper bound on the EDM of the neutron is $(d_n/e) < 
1.1\ 10^{-25}$ cm, at 95$\%$ of confidence level (CL).\cite{PDG} 
However, taking the typical values of $m_d=10$ MeV and $M_W\approx M\approx 
100$ GeV, Eq.\ (\ref{EDMf}) gives $(d_n/e) < 3.\ 10^{-30}\, \xi_{HA}$ cm, 
far beyond the above experimental bound. The prediction for the EDM
of electron is less restrictive. On the other hand, the two-loop Barr-Zee
(BZ) mechanism\cite{BZ} shown in Fig.\ 1 may have a significant contribution
to the EDM, leading to tighter bounds on the $HA$ mixing parameter $\xi_{HA}$.
In general, the theoretic prediction for the EDM is enhanced by a factor 
$\alpha_{em} (M^2_W/m^2_d) \approx 10^6$, with $\alpha_{em} = 1/137$. 
A recent analysis of constraints from the electron and neutron EDM's may be
found in Ref.\cite{EDMnew} for the two-Higgs doublet model with maximal CP 
violation, in which the Weinberg's unitarity bound is almost 
saturated.\cite{SW}  The authors\cite{EDMnew} find that the bounds on the 
neutron EDM may be evaded if the mass difference, $\Delta M$, between $A$ 
and $H$ is sufficiently small, {\em viz.} 
\begin{equation}
\label{EDMHA}
\frac{\Delta M}{M}\ \approx\ \frac{\widehat{\Pi}^{HA}}{M^2}\ <\ 0.10,\
0.13,\ 0.24\, ,
\end{equation}
for $M=200,\ 400,\ 600$ GeV, respectively. Since we always have 
$\widehat{\Pi}^{HA}/M^2<0.1$ in our two-Higgs doublet models,
the EDM limits in Eq.\ (\ref{EDMHA}) are therefore satisfied. 
For large $\tan\beta$ values, far way from the parametric point of 
maximal CP violation, the above EDM limits will be much weaker. However, 
our resonant CP-violating phenomena through particle mixing can still be 
very large as soon as the necessary conditions (\ref{CPN1}) and/or 
(\ref{CPN2}) are fulfilled.

\section{CP violation through $HA$ mixing at high-energy colliders}

The most ideal place to look for resonant CP-violating $HA$
transitions is at $e^+e^-$ and, most interestingly, at muon 
colliders.\cite{mumu} In general, there are many observables suggested at
high-energy colliders\cite{DV,NP1,PN,Gavela,Nacht,Peskin,WB,CK,CPmumu,RCPV} 
that may be formed to project out different CP/T-noninvariant
contributions. All the CP-violating observables, however, may fall
into two categories, depending on whether they are even or odd under
naive CPT transformations.  Our quantitative analysis of $HA$ mixing
phenomena will rely on CP-violating quantities sensitive to CP- and
CPT-odd observables of the form, $\langle \vec{s}_t \vec{k}_t \rangle$
or $\langle \vec{s}_{\bar{t}} \vec{k}_{\bar{t}}\rangle$,\cite{PN,Peskin,CK} 
where $\vec{s}_t$ and $\vec{k}_t$ are respectively the spin and the 
three-momentum of the top-quark in the $t\bar{t}$ centre of mass (c.m.) 
system.  For definiteness, assuming that having longitudinally polarized 
muon beams will be feasible without much loss of luminosity, we shall consider
the CP asymmetry\cite{CPmumu}
\begin{equation}
\label{CPmuobs}
{\cal A}^{(\mu)}_{CP}\ =\ \frac{\sigma (\mu^-_L\mu^+_L\to f\bar{f})\
-\ \sigma (\mu^-_R\mu^+_R\to f\bar{f})}{\sigma (\mu^-_L\mu^+_L\to
f\bar{f})\ +\ \sigma (\mu^-_R\mu^+_R\to f\bar{f})}\ .
\end{equation}
If one is able to tag on the final fermion pair $f\bar{f}$ ({\em
e.g.}, $\tau^+\tau^-$, $b\bar{b}$, or $t\bar{t}$), ${\cal A}^{(\mu
)}_{CP}$ is then a genuine observable of CP violation, because the helicity
states $\mu^-_L\mu^+_L$ transform into $\mu^-_R\mu^+_R$ under CP in
the c.m.\ system. Similarly, at $e^+e^-$ or $p\bar{p}$ machines, one
can define the CP asymmetry
\begin{equation}
\label{CPelobs}
{\cal A}^{(e)}_{CP}\ =\ \frac{\sigma (e^-e^+\to f_L\bar{f}_L\, X)\ -\
\sigma (e^-e^+\to f_R\bar{f}_R\, X)}{\sigma (e^-e^+\to f_L\bar{f}_L\,
X)\ +\ \sigma (e^-e^+\to f_R\bar{f}_R\, X)}\ ,
\end{equation}
and an analogous observable ${\cal A}^{(p)}_{CP}$. In Eq.\
(\ref{CPelobs}), the chirality of fermions, such as the top quark, may
not be directly observed. However, the decay characteristics of a
left-handed top quark differ substantially from those of its
right-handed component, giving rise to distinct angular-momentum
distributions and energy asymmetries of the produced charged leptons
and jets.\cite{Peskin,WB}

A CP-conserving  two-Higgs doublet model predicts three physical Higgs
scalars, from which two are CP even and one is CP odd. To break the
CP invariance of the Higgs sector, one may have to introduce soft D-symmetry
breaking terms in the Higgs potential, which violate CP as well. In this
way, one does not spoil the desirable property of the imposed D symmetry, 
{\em i.e.}, flavour-changing neutral currents are induced beyond tree level.
Another alternative is to break CP invariance radiatively after integrating
out heavy degrees of freedom which do not respect CP. As such, one may
think of heavy Majorana fermions, such as heavy Majorana neutrinos or
neutralinos in supersymmetric models. Here, we adopt the former realization
and fix the three heavy neutrino masses to be $m_1=0.5$ TeV, $m_2=1$ TeV and 
$m_3=1.5$ TeV. In the two-Higgs doublet model with broken D symmetry, 
the tree-level $HA$ or $hA$ mixings, $\Pi^{HA}$ and $\Pi^{hA}$, are considered
to be small phenomenological parameters in compliance with the constraint of 
Eq.\ (\ref{EDMHA}). For the top quark mass, we use $m_t=170$ GeV close to its
experimental mean value.\cite{PDG}

For simplicity, we shall assume that only one CP-even Higgs particle,
$H$ say, has a mass quite close to $A$, {\em i.e.}, $M_H - M_A \ll
M_H, M_A$, while the other CP-even Higgs, $h$, is much lighter than
$H$, $A$, and vice versa.  Clearly, $h$ will effectively decouple from
the mixing system, having negligible contributions to both cross
section and CP asymmetry at c.m.\ energies $s\approx M_H,M_A$. This is
also the main reason accounting for the fact that CP violation through
$HZ$ ($G^0H$) mixing has been found to be small,\cite{APRL} as $H$
only couples to the longitudinal component of the $Z$ boson, the
massless would-be Goldstone $G^0$. To give an estimate, in the
two-Higgs model with heavy Majorana neutrinos, we find that ${\cal
A}^{(\mu )}_{CP}\approx 2.\ 10^{-2}$ for $M_H=500$ GeV, while the 
production cross-section is $\sigma \simeq 1$ fb. It is therefore unlikely 
to observe $HZ$-mixing effects, even if one assumes a high integrated 
luminosity of 50 fb$^{-1}$, designed for $e^+e^-$ and $\mu^+\mu^-$ colliders. 
For our illustrations, we also take the ratio of the vacuum expectation values
of the two Higgs doublets $\tan\beta=2$ and $\tan\theta=1$, which
relates the weak with the mass eigenstates for the CP-conserving Higgs
particles, $H$ and $h$. In this scheme, the heaviest CP-even Higgs,
$H$, has a significant coupling to fermions, whereas the squared of
the $HWW$- and $HZZ$-couplings has strength 1/10 of that of the
respective SM couplings.  For $M_A > 2 M_Z$, such a scheme is
motivated by the MSSM and leads to nearly degenerate $H$ and $A$
scalars, {\em i.e.}, $M_H \approx M_A$.\cite{JFG} Nevertheless, we
shall not consider here that $M_H$ is very strongly correlated with
$M_A$.

We shall now focus our attention on the resonant transition amplitudes
$\mu^+_L\mu^-_L \to H^*,\, A^* \to f\bar{f}$. A straightforward calculation 
of the CP asymmetry gives
\begin{equation}
\label{ACPmu}
{\cal A}^{(\mu )}_{CP}(s)\ \approx\ 
\frac{ 2\, \widehat{\Pi}^{AH}\, (\beta_f \Im m\widehat{\Pi}^{AA} \,
- \, \Im m\widehat{\Pi}^{HH} ) }{ \beta_f [(s-M^2_A)^2 + (\Im
m\widehat{\Pi}^{AA})^2] + (s-M^2_H)^2 + (\Im m\widehat{\Pi}^{HH})^2}\ ,
\end{equation}
where $\beta_f = (1-4m^2_f/s)$ with $r_f=\chi^f_H/\chi^f_A$.  
The analytic result of ${\cal A}^{(\mu )}_{CP}(s)$ in Eq.\
(\ref{ACPmu}) reduces to the qualitative estimate presented in Ref.\cite{APRL},
if finite mass effects of the asymptotic states are neglected. Refinements of 
including high-order $(\widehat{\Pi}^{AH})^2$ terms in the CP-conserving part 
of the CP asymmetry and other model-dependent details inherent to the two-Higgs
doublet model may be found in Ref.\cite{RCPV}

Assuming that tuning the collider c.m.\ energy to the mass of $H$ or
$h$ is possible, {\em i.e.}, $\sqrt{s}=M_H$ or $M_h$, we analyze the
following two reactions:
\begin{eqnarray}
\mbox{(a)} && \mu^+_L\mu^-_L \to h^*,A^* \to b\bar{b}\, ,\quad
\mbox{with}\ M_A=170\ \mbox{GeV}\, ,\nonumber\\ \mbox{(b)} &&
\mu^+_L\mu^-_L \to H^*,A^*\to t\bar{t}\, ,\quad \mbox{with}\ M_A=400\
\mbox{GeV}\, . \nonumber
\end{eqnarray}
As has been discussed in Section 2, CP violation becomes maximal when
the necessary condition (\ref{CPN1}) for $M_H\approx M_A$ is met. In
Fig.\ 2(a), we show how $|A_{CP}|$ varies as a function of the
parameter $x_A=\Im m\Pi^{HA}/\Im m(\widehat{\Pi}^{HH}-\widehat{\Pi}^{AA})$ 
or $\Im m\Pi^{hA}/\Im m(\widehat{\Pi}^{HH}-\widehat{\Pi}^{AA})$. We consider 
the kinematic region for resonantly enhanced CP violation, {\em i.e.}, 
$M_A=M_h$ ($M_H$) for the reaction (a) (reaction (b)). We find that 
CP-violating effects could become very large, if the parameter $x_A$ was 
tuned to the value $x_A=1$ for the process (a) and $x_A=3$ for the process (b).
  
\begin{figure}[ht]
   \leavevmode
 \begin{center}
   \epsfxsize=15.0cm \epsffile[0 0 539 652]{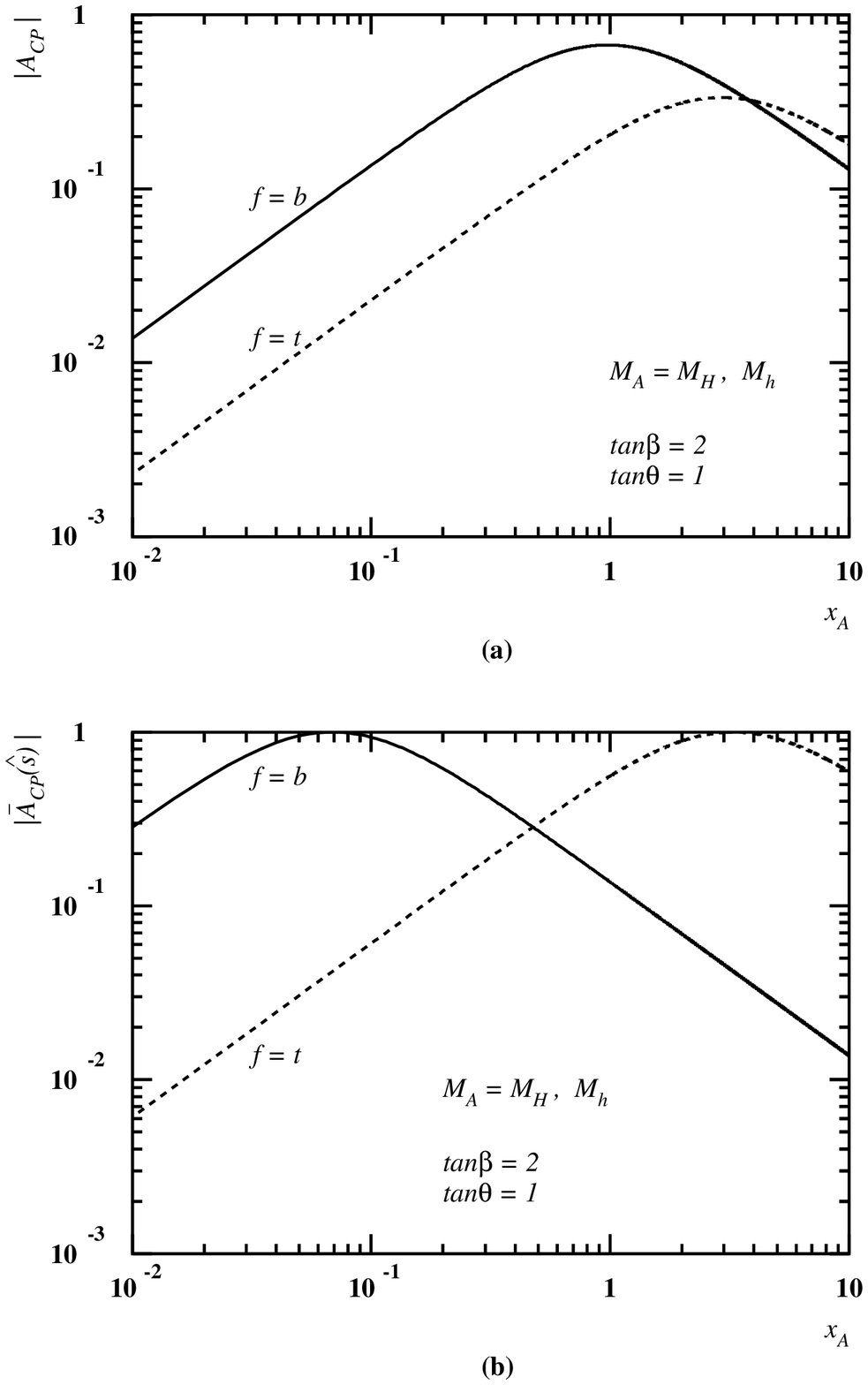}
{\small {\bf Fig.\ 2:} CP asymmetries at (a) the muon collider and (b) the NLC as a function of $x_A$.}
 \end{center}
\end{figure}

At future  next linear $e^+e^-$ colliders (NLC), Higgs bosons can
copiously be produced either via the Bjorken process for c.m.\ energies
up to 0.5 TeV or through $WW$ fusion at higher energies.\cite{NLC} The
most convenient way is to study CP violation in the kinematic range of
the Higgs production and decay.\cite{PN,BG} Therefore, we
shall be interested in the observable
\begin{equation}
\label{CPeldif}
\overline{{\cal A}}^{(e)}_{CP}(\hat{s})\ =\ \frac{d\sigma (e^-e^+\to
f_L\bar{f}_L\, X)/d\hat{s}\ -\ d\sigma (e^-e^+\to f_R\bar{f}_R\,
X)/d\hat{s}}{ d\sigma (e^-e^+\to f_L\bar{f}_L\, X)/d\hat{s}\ +\
d\sigma (e^-e^+\to f_R\bar{f}_R\, X)/d\hat{s}}\ ,
\end{equation}
where $\hat{s}$ is the invariant mass energy of the produced final
fermions $f$. As final states, we may take bottom or top quarks.
Since $A$ does not couple to $WW$ or $ZZ$, the CP asymmetry
$\hat{{\cal A}}^{(e)}_{CP}$ in Eq.\ (\ref{CPeldif}) takes the simple
form
\begin{equation}
\label{ACPel}
\overline{{\cal A}}^{(e)}_{CP}(\hat{s}) \ \approx\ -\, \frac{
2\widehat{\Pi}^{AH}(\hat{s})\, \Im m\widehat{\Pi}^{AA}(\hat{s})}{
(\hat{s}-M^2_A)^2 + (\Im m\widehat{\Pi}^{AA}(\hat{s}))^2 +
(\widehat{\Pi}^{AH}(\hat{s}))^2}\, .
\end{equation}
Since the destructive term, $\Im m\widehat{\Pi}^{HH}$, is absent in
Eq.\ (\ref{ACPel}), CP violation may become even larger, {\em i.e.},
of order unity for specific values of the parameter $x_A$. Indeed, we
see from Fig.\ 2(b) that $\overline{{\cal A}}^{(e)}\approx 1$, if
$x_A=0.07$ (3) for the reaction with longitudinally polarized $b$
($t$) quarks in the final state.  In Ref.,\cite{PN} it has been estimated
numerically that ${\cal A}^{(e)}_{CP} < 15\%$ for Higgs masses
$M_H<600$ GeV and cross sections $\sigma (e^-e^+\to H^* (h^*), A^*\to
t\bar{t}\, X)\approx 10-100$ fb for c.m.\ energy of 2 TeV.  Such
CP-violating effects have good chances to be detected at the NLC.

At the LHC, the respective CP asymmetry $\overline{{\cal A }}^{(p)}_{CP} 
(\hat{s})$ may be obtained from Eq.\ (\ref{ACPel}), for Higgs
particles that have a production mechanism similar to that at the NLC.
If the Higgs is produced via gluon fusion, one has to use the analytic
expression of ${\cal A}^{(\mu )}_{CP}(\hat{s})$ in Eq.\ (\ref{ACPmu}).
Unless CP violation is resonantly amplified, {\em i.e.}, of order one,
the chances to detect CP-violating phenomena on the Higgs-resonance
line after removing the contributing background appear to be quite
limited at the LHC. It is therefore worth stressing that a large
CP-violating signal at the Higgs-boson peak will certainly point
towards the existence of an almost degenerate $HA$ mixing system.

\section{Conclusions}

CP-violating phenomena can be significantly enhanced through the mixing of 
two resonant particles that behave differently under CP. In particular, the 
underlying mechanism for large CP violation induced by resonant $HA$
transitions has been studied carefully on a rigorous field-theoretic 
basis and its connection with the $K^0\bar{K}^0$ system has been clarified
in Section 2. Possible constraints resulting from experimental bounds on 
the neutron EDM are briefly discussed in Section 3. In Section 4, the size 
of CP-violation in the production, mixing and decay of a Higgs particle at 
planned high-energy machines, such as the LHC, NLC and/or muon collider, has 
been estimated. Since high order $\varepsilon'$-type effects are generally 
suppressed near the resonant region, possible large CP-violating phenomena 
can naturally be accounted for by the mixing mechanism presented in this talk.

\end{document}